# Controlling phonon emission with plasmonic metamaterials

K. Kempa

*Department of Physics, Boston College.*

**Abstract**

It is shown, that plasmonic metamaterial nanostructures could be used to reduce the electron-phonon scattering rate, by providing an alternative, fast electron-plasmon scattering channel. Since the plasmon-phonon and plasmon-photon scattering processes are relatively slow, this provides a mechanism for a hot-electron plasmonic protection against the phonon emission. The stored/protected energy can be returned to the single particle channel by processes similar to the Rabi oscillations, plasmon resonance energy transfer (PERT), or formation of plasmarons. This effect could be used to control phonon scattering in various electron systems, such as solar cells or high Tc-superconductors.

Electron-phonon scattering is a ubiquitous phenomenon in all condensed matter systems, and leads to a rapid thermalization of excited electrons to the lattice temperature. Controlling the electron-phonon scattering has been a fundamental challenge in physics, and has been proven very difficult. For example, the quest for the increased critical temperature in conventional superconductors included attempts to engineer the phonon spectrum to enhance the electron-phonon coupling [1]. Recently, it has become clear, that increasing the critical temperature in the high-Tc materials will most likely involve an electron-phonon engineering, this time to reduce the deleterious carrier scattering by high energy phonons [2]. In another example, the solar cells could be dramatically improved, if phonon losses of the hot electrons excited well above the conduction band edge by high-energy photons of the solar spectrum, could be prevented. A typical, high performance crystalline silicon solar cell has the energy conversion efficiency of only about 20%, while almost 30% of the solar energy is used in this solar cell to generate heat (phonons) [3].

The problem with controlling electron-phonon scattering is that this is a very fast process, involving a very rich spectrum of phonon excitations. Attempts to control this process have been partially successful in quantum dots, were the so called "phonon bottleneck" was demonstrated [4], and in thermoelectric materials and structures, where a decoupling of the phonon and electron channels is achieved by a superlattice structuring, or nanoparticle composites [5]. The only process that could compete directly (have equal or larger scattering rate) with the phonon emission by an excited electron is the plasmon scattering [6], provided that the electron density is high enough, like that in metals or highly doped semiconductors. Following this fact, a scheme is proposed here, in which a hot electron in an open band (say of a semiconductor) is coupled to a metallic plasmonic metamaterial nanostructure (PMN), which sustains robust plasmon oscillations in a proper frequency range. This way the electron could emit a plasmon instead a phonon, as long as the plasmon scattering rate was high enough. In contrast to phonons, plasmons can couple directly back to the single particle channel in the open band, thereby restoring at least part of the stored energy to electrons.

In general, the scattering rate of an excited electron from the state $E_\mathbf{k}$ to states $E_\mathbf{k+q}$, due to single particle and collective (plasmon) excitations (with wave vectors $\mathbf{q}$) is given by [6-9]

$$\gamma_\mathbf{k} \approx -\frac{2}{\hbar}\int \frac{d\mathbf{q}}{(2\pi)^3} v_q \left[ n_B(E_\mathbf{k} - E_\mathbf{k+q}) - n_F(-E_\mathbf{k+q} + \mu) \right] \mathrm{Im}\left[ V_{eff}[q,(E_\mathbf{k+q} - E_\mathbf{k})/\hbar] \right] \quad (1)$$

where $n_B$ and $n_F$ are the Bose-Einstein and Fermi-Dirac distribution functions correspondingly, $\mu$ is the chemical potential, $v_q$ is the bare Coulomb interaction, and $V_{eff}(q,\omega)$ is the dressed combined interaction, which can be written as a simple sum of the Coulomb and phonon (Frohlich) terms [6,8]

$$V_{eff}(q,\omega) = \frac{V_q}{\varepsilon(q,\omega)} + \frac{\Omega_q |g_q/\varepsilon(q,\omega)|^2}{\omega^2 - \Omega_q^2/\varepsilon(q,\omega) + i0^+} \quad (2)$$

where $\varepsilon(q,\omega)$ is the longitudinal dielectric function of the medium, $g_q$ is the matrix element, $\Omega_q$ is the longitudinal phonon frequency (plasma frequency of the ionic "plasma"). Eq. (2) is written in the random phase approximation (RPA) for electrons (first term), and the point-ion, long wavelength approximation for ions (second term) [6,8].

From Eq. (2) it is clear, that the electron scattering is controlled by that with other electrons (first, Coulomb term), and phonons (second, Frohlich term). It is also clear, that the strongest contribution to the electron-electron scattering comes from the collective branch (plasmons), for which $\varepsilon(q,\omega)$ vanishes. In this work, the focus is on calculating the scattering of an excited electron in a medium (e.g. semiconductor) with plasmons in a PMN, embedded or strongly coupled to the medium. This is subsequently compared to the electron-phonon, and other scattering mechanisms. The calculations of the electron-plasmon scattering is facilitated by simplicity of the dielectric function of PMN-semiconductor system, which in the effective medium limit [10-12], can be well described by an effective, local dielectric function of the general form [13-16]

$$\varepsilon(\omega) = \varepsilon_{back} + \sum_{f=1}^{M} \frac{\omega_{pf}^2}{\omega_{rf}^2 - \omega^2} \tag{3}$$

Here $\omega$ is the frequency of the electromagnetic radiation, and $\omega_{rf}, \omega_{pf}$, and $\varepsilon_{back}$ are constants. For PMN in the form of a 2D island array $\omega_{r1} \neq 0$, but for a 2D hole array $\omega_{r1}$ must vanish. Bulk, longitudinal plasmons occur anytime $\varepsilon(\omega) = 0$. The simplest form of Eq. (2), which includes propagating and trapped plasmon modes is

$$\varepsilon(\omega) = \varepsilon_b + \frac{\omega_p^2}{\omega_r^2 - \omega^2} \tag{4}$$

This form is also an exact effective dielectric function for the 3D point-dipole crystal [17], and (with $\omega_r = 0$) was used to describe the extraordinary optical transmission (EOT) [19] of nanoscopically perforated metallic films in the subwavelength limit [19,20]. With $\varepsilon(q,\omega)$ given by Eq. (4), and at room temperatures one obtains from Eq. (1) and Eq. (2) with the Coulomb term only, the following, explicit expression for the electron-plasmon scattering rate in our system

$$\gamma_{el-pl} \approx \frac{\sqrt{2E_k/m^*}}{2a_0^*} f\left(\frac{E_k}{E_0}\right)\theta\left(\frac{E_k}{E_0}-1\right) \qquad (5)$$

where

$$a_0^* = \hbar^2 \varepsilon_b / m^* e^2 \qquad (6)$$

$$E_0 = \frac{\hbar\omega_p}{\sqrt{\varepsilon_b}} \left[\varepsilon_b \sqrt{1+\varepsilon_b (\omega_r/\omega_p)^2}\right]^{-1} \qquad (7)$$

$$f(x) = \frac{2}{x} \ln\left[\sqrt{x}+\sqrt{x-1}\right] \qquad (8)$$

In the limit of $\varepsilon_b = 1$ and $\omega_r = 0$ (for which $E_0 = \hbar\omega_p$), Eq. (5) reduces to the well-known form for bulk metals [6]. It was shown, that with a more realistic model of the metallic dielectric response (e.g. the random phase approximation, RPA), the results are only marginally modified (to within ~ 10%), and that these results are in agreement with experiment [6]. The universal auxiliary function, given by Eq. (8), which controls the $E_k$ dependency of $\gamma_{el-pl}$, is plotted in Fig. 1. As expected, it shows that the electron-plasmon scattering occurs for $E_k > E_0$, an effective plasmon frequency of PMN, and the rate sharply picks at about 1.7 $E_0$, emphasizing the known fact, that a few plasmon scattering is preferential [6].

The calculation of the electron-plasmon scattering rate $\gamma_{el-pl}$, obtained from Eq. (5) with parameters for crystalline silicon and PMN designed so that $E_0 = 0.25$ eV, for various values of the initial electron energy $E_k$ is shown in Fig. 2 (bold solid line). The rate is very high, and well exceeds in almost the entire range of $E_k$ the electron-phonon scattering rate $\gamma_{el-ph}$, represented by the vertical bars. $\gamma_{el-ph}$ was simulated elsewhere by various methods [21-25], including *ab initio* calculation [23], which yields electron mobilities in excellent agreement with experiment [25].

Note, that this conclusion is unchanged, if instead of one, there are many electrons excited in the semiconductor band. Then both rates ($\gamma_{el-pl}$ and $\gamma_{el-ph}$) are getting smaller, as a result of a shrinking momentum space for available transitions, but essentially by the same factor, so that their ratio is preserved. Even though, the electron-plasmon scattering "outperforms" the electron phonon scattering, use of the PMN as a plasmonic reservoir, "protecting" the electron energy from phonon dissipation, rests on stability of this reservoir against the phonon and radiative losses, i.e. we must show, that the plasmon-phonon and plasmon-photon scattering rates in PMN are much smaller than the electron-phonon scattering in the semiconductor, i.e. $\gamma_{pl-ph} \ll \gamma_{el-ph}$ and $\gamma_{pl-phot} \ll \gamma_{el-ph}$.

To estimate $\gamma_{pl-ph}$, we first notice that dispersion of any plasmonic (or polaritonic) mode is given in general by

$$F[\varepsilon(q,\omega)] = 0 \tag{9}$$

where $F[x]$ is an analytic function of $x$. Let assume, that Eq. (9) has the following solution $\omega = \omega_0(q)$. A general way to account for losses in expressions for the dielectric functions (of the form Eq. (4)) is to replace $\omega^2$ with $\omega(\omega + i\gamma)$, where $\gamma$ is the rate of inelastic scattering with the lattice (essentially an average of $\gamma_{el-ph}^{\mathbf{k}}$). Parameter $\gamma$ is known experimentally for most metals. Now, consider the following expression

$$\bar{\omega} = \sqrt{\omega_0^2(q) - \frac{\gamma^2}{4}} - i\frac{\gamma}{2} \tag{10}$$

Since, $\bar{\omega}(\bar{\omega} + i\gamma) = \omega_0^2(q)$, and Eq. (3) contains only $\omega^2$ (replaced with $\omega(\omega + i\gamma)$), this implies that $\omega = \bar{\omega}$ is also a solution to Eq. (9), and thus any plasmonic (polaritonic) mode scatters with phonons at an average rate of $\gamma/2$. Since for silver (best plasmonic metal) $\gamma \approx 0.25 \times 10^{14} (\text{sec})^{-1}$ [26], we find $\gamma_{pl-ph} \approx \gamma/2 \approx 0.125 \times 10^{14} (\text{sec})^{-1} \ll \gamma_{el-ph}$. $\gamma_{pl-ph}$ is represented in Fig. 2 as a thin-horizontal line. This estimated plasmon-phonon scattering rate agrees well with the Mie plasmon peak broadening due to the Drude damping, as calculated and measured in Ref. [27] for PMN consisting of a planar, periodic array of ellipsoidal Ag nanopartitles (80 nm x 40 nm) with the lattice period of 200 nm.

To estimate $\gamma_{pl-phot}$, we note that this radiative damping scales as $\omega^4$ [28], and thus it is not expected to be important at the IR frequencies. This is fully confirmed by detailed calculations in Ref. [27], which show that while $\gamma_{pl-phot} > \gamma$ in the visible frequency range, it is only $\gamma_{pl-phot} \approx 10^{11} (\text{sec})^{-1}$ at the frequency of the plasmon resonance $E_0 = 0.25$ eV, and thus $\gamma_{pl-phot} \ll \gamma$. Therefore, this plasmon-photon scattering process can be here ignored. With these results, Fig. 2 clearly demonstrates, that electron-plasmon scattering process is faster than the electron-phonon process in the semiconductor, and much faster than any scattering processes in PMN in the entire relevant energy range. This is the condition for the plasmonic protection; once the hot electron energy is transferred to plasmons in PMN, it remains there "protected" from all relevant emissions (phonons and photons). *This is the main result of this work.*

The remaining, important issue is the recovery of the stored energy from the plasmonic reservoir. A possible mechanism could be similar to the plasmon resonance energy transfer (PERT), observed between metallic and semiconducting nanoparticles [29]. Since in our scheme PMN is strongly coupled to the excited electron/electrons in the semiconductor, this could transform the excited plasmon into a plasmaron, a coupled plasmon-single particle excitation. Such an excitation has been predicted by Lundqvist [30], and later have been observed in elemental bismuth [31], and most recently in graphene [32]. In addition, the plasmonic/plasmaronic resonator (PMN) acts as a high Q resonator of the electromagnetic field. Thus, the conditions can arise for Rabi-like oscillations, in which energy of a hot electron oscillates between the electron and the plasmonic/plasmaronic reservoir. The period of these oscillations is expected to be proportional to the matrix element involving the initial and final states of the hot electron, and the electric field of the reservoir (PMN structure). In fact, such Rabi-like coupling mechanism has been already observed in a PMN structure coupled to semiconductor quantum dots [33]. This could be a very efficient energy recovery mechanism, provided it is optimized for a specific application.

Finally, I comment on a possible structure design, which could directly benefit from this idea. Recently, a high efficiency solar cell was proposed, based on a plasmonic metamaterial design [34]. This structure consists of an ultra-thin amorphous silicon absorber film (*p-i-n* junction), sandwiched between two metallic films, one continuous, and the other in form of a PMN, both made of silver. The structure was shown to act as an excellent broad-band absorber, if PMN was chosen to be a checkerboard structure [34]. It might be possible to augment the excellent absorption capability of this structure, with the plasmonic protection. In one scenario, one could engineer one of the two metallic layers in this structure (by proper texturing), so that it functions as a plasmonic resonator with the desired ($E_0$). In another scenario, an additional PMN could be added to the structure, or an array of nanoparticles (either 2D or 3D) embedded directly into the absorber. In either case, designing such a complex structure will require quantitative simulations (e.g. FDTD) of the optical performance, as well as calculations/simulations (RPA or TDLDA) of the carrier dynamics.

In conclusion, it was shown, that a plasmonic metamaterial metallic nanostructure could be used to reduce the electron-phonon scattering rate, by providing an alternative, fast electron-plasmon scattering channel. Since the plasmon-phonon and plasmon-photon scattering processes are relatively slow in this structure, this could provide a plasmonic protection mechanism for the excess free energy of the excited (hot) electrons, and in principle also holes. This effect could help to control the electron-phonon scattering in various systems, and could be useful in various applications, including solar cells.

**Acknowledgements**

I would like to thank Prof. M.J. Naughton for discussions and for suggesting connection to PERT.

**Figure Captions**

**Fig. 1**. The universal auxiliary function $f(x)$ given by Eq. (8), plotted versus $x = E_k / E_0$.

**Fig. 2**. Electron scattering rates of hot electrons in silicon. Scattering with plasmons in PMN (calculated from Eq. (5), bold-solid line), with phonons in silicon (simulations done elsewhere, vertical bars). The horizontal thin-solid line represents the plasmon-phonon scattering rate in PMN, estimated from Eq. (10). All plotted versus the electron energy $E_k$.

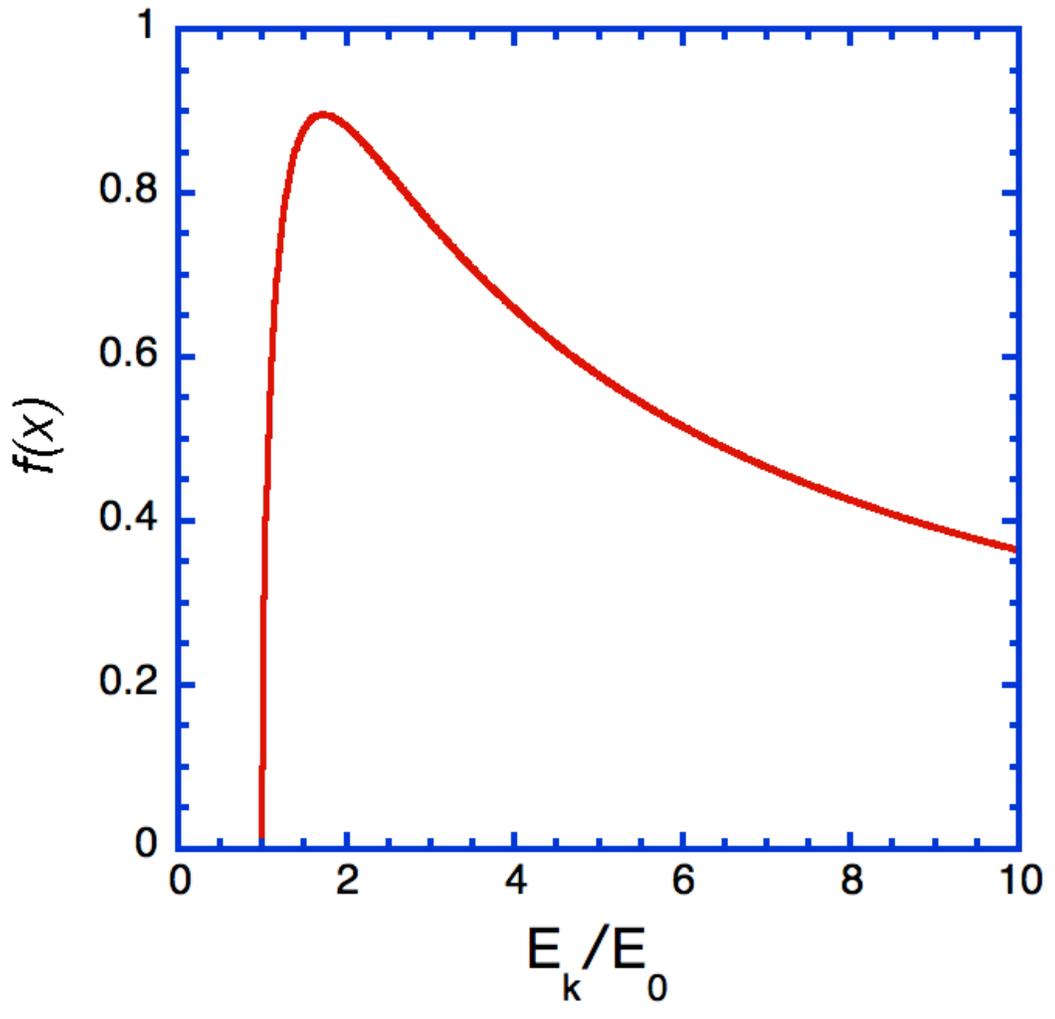

Fig. 1

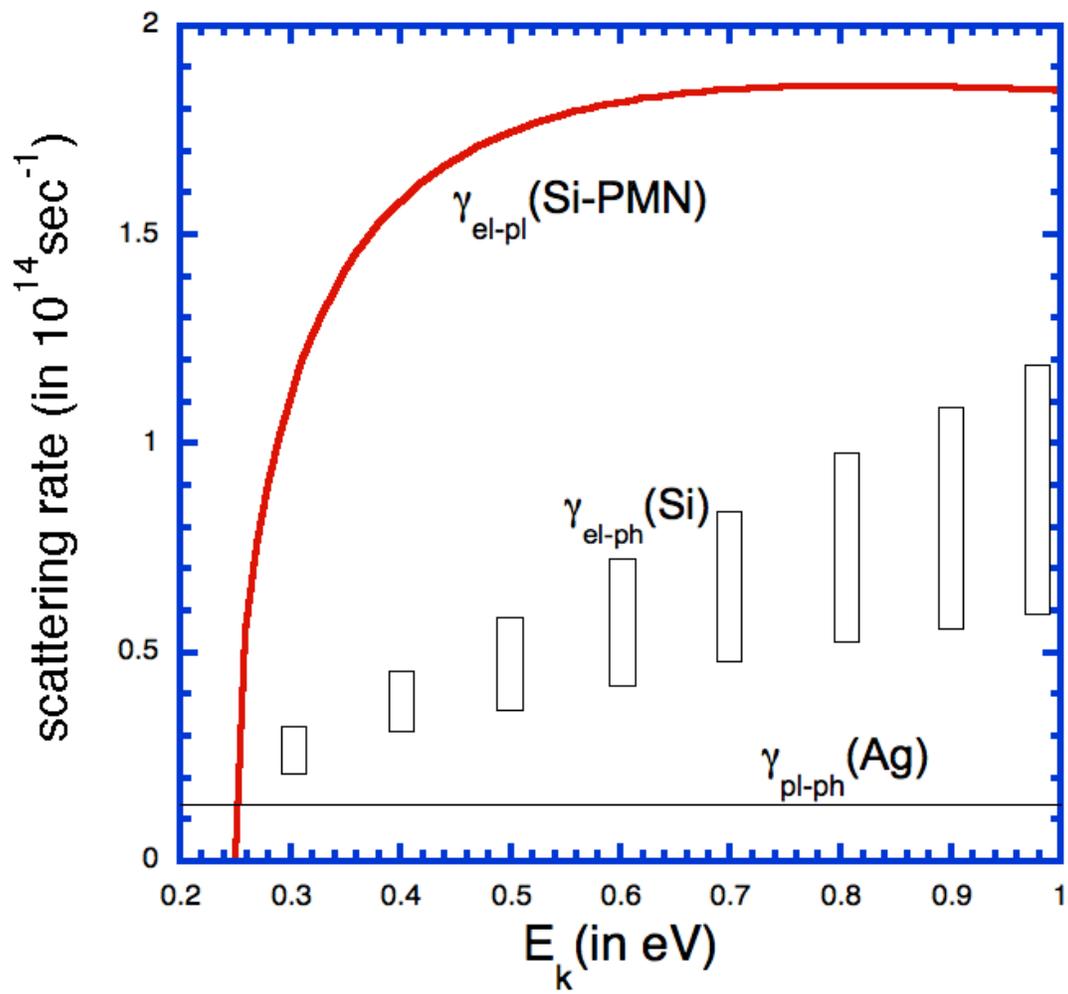

Fig. 2